\documentclass{emulateapj}
\usepackage[utf8]{inputenc}
\usepackage[T1]{fontenc}
\usepackage{graphicx}
\usepackage{color}
\usepackage[dvipsnames]{xcolor}
\usepackage{amsmath}

\newcommand{\Msun}{\ensuremath{\,M_\odot}}
\newcommand{\Rsun}{\ensuremath{\,R_\odot}}
\newcommand{\Zsun}{\ensuremath{\,Z_\odot}}
\newcommand{\Lsun}{\ensuremath{\,L_\odot}}

\newcommand{\yr}{\ensuremath{\,\mathrm{yr}}}
\newcommand{\kyr}{\ensuremath{\,\mathrm{kyr}}}
\newcommand{\Myr}{\ensuremath{\,\mathrm{Myr}}}
\newcommand{\Gyr}{\ensuremath{\,\mathrm{Gyr}}}

\newcommand{\kms}{\ensuremath{\,\mathrm{km}\,\mathrm{s}^{-1}}}
\newcommand{\ergs}{\ensuremath{\,\mathrm{erg}\,\mathrm{s}^{-1}}}
\newcommand{\msy}{\ensuremath{\Msun\mathrm{\; yr}^{-1}}}

\newcommand{\mpc}{\ensuremath{\,\mathrm{Mpc}}}

\newcommand{\Ledd}{\ensuremath{L_{\rm Edd}}}
\newcommand{\Medd}{\ensuremath{\dot M_{\rm Edd}}}

\newcommand{\startrack}{{\tt StarTrack}}

\newcommand{\sci}[2]{\ensuremath{#1\times10^{#2}}}

\begin{document}

\title{Wind-powered ultraluminous X-ray sources}

\author{Grzegorz Wiktorowicz\altaffilmark{1,2,3}\thanks{E-mail: gwiktoro@astrouw.edu.pl},
        Jean-Pierre Lasota\altaffilmark{3,4},
        Krzysztof Belczynski\altaffilmark{3},
        Youjun Lu\altaffilmark{1,2},
        Jifeng Liu\altaffilmark{1,2,5},
        Krystian Iłkiewicz\altaffilmark{3,6,7}}

 \affil{  
     $^{1}$ National Astronomical Observatories, Chinese Academy of Sciences, Beijing 100101, China\\
     $^{2}$ School of Astronomy \& Space Science, University of the Chinese Academy of Sciences, Beijing 100012, China\\
     $^3$ Nicolaus Copernicus Astronomical Center, Polish Academy of Sciences, Bartycka 18, 00-716 Warsaw, Poland\\
    $^4$ Institut d’Astrophysique de Paris, CNRS et Sorbonne Université, UMR 7095, 98bis Bd Arago, 75014 Paris, France\\
    $^{5}$ WHU-NAOC Joint Center for Astronomy, Wuhan University, Wuhan, China\\
    $^{6}$ Centre for Extragalactic Astronomy, Department of Physics, University of Durham, South Road, Durham, DH1 3LE, UK\\
    $^{7}$ Department of Physics and Astronomy, Box 41051, Science Building, Texas Tech University, Lubbock, TX 79409-1051, USA\\
}
 
\begin{abstract}

Although ultraluminous X-ray sources (ULX) are important for astrophysics because of their extreme apparent super-Eddington luminosities, their nature is still poorly known. Theoretical and observational studies suggest that ULXs could be a diversified group of objects that are composed of low-mass X-ray binaries, high-mass X-ray binaries and marginally also systems containing intermediate-mass black holes. Observational data on the ULX donors could significantly boost our understanding of these systems, but only a few have been detected. There are several candidates, mostly red supergiants (RSGs), but surveys are typically biased toward luminous near-infrared objects. In ULXs harbouring RSGs matter accreted onto the compact body would have to be provided by the stellar wind of the companion because a Roche-lobe overflow could be unstable for relevant mass-ratios. Here we present a comprehensive study of the evolution and population of wind-fed ULXs, and we provide a theoretical support for the link between RSGs and ULXs. Assuming a minimal model of stellar-wind emission, our estimated upper limit on contribution of wind-fed ULX to the overall ULX population is $\sim75$--$96\%$ for young ($<100\Myr$) star forming environments, $\sim 49$--$87\%$ for prolonged constant star formation (e.g., disk of Milky Way), and $\lesssim1\%$ for environments in which star formation ceased long time ($>2\Gyr$) ago. We demonstrate that some wind-fed ULXs (up to $6\%$) may evolve into merging double compact objects (DCOs). We demonstrate that, the exclusion of wind-fed ULXs from population studies of ULXs might have lead to systematic errors in their conclusions.

\end{abstract}


\section{Introduction}

Ultraluminous X-ray sources (ULXs) are defined as point-like off-nuclear X-ray sources with luminosities exceeding $10^{39}\ergs$ \citep[for a recent review see][]{Kaaret1708}. Therefore, they are the brightest X-ray sources that are neither supernovae (SNe) nor active galactic nuclei. After their identification at the end of the previous century, it was thought for a long time that ULXs contain intermediate-mass ($M \sim 10^2$--$10^4\Msun$) black holes \citep[IMBHs][]{Colbert9907} that are accreting at rates close to or below their Eddington values. The dissenting view of \citet{King0105}--who pointed out that there are no viable evolutionary scenarios leading to the formation of such systems and suggested that most ULXs are stellar-mass binaries that are accreting at super-Eddington rates--was largely ignored until \citet{Bachetti1410} discovered that the source M82 ULX-2 contains an X-ray pulsar. This observation and the subsequent observations of five other pulsating ULXs (PULXs) by \citet{Israel1609}, \citet{Furst1701}, \citet{Carpano1805}, \citet{Sathyaprakash1909}, \citet{Rodriguez-Castillo2005} led to a change of paradigm and it is now universally believed that most (if not all) ULXs contain stellar-mass accretors. PULXs have been also observed during giant outbursts of Be-X-ray binary stars \citep[one in the Galaxy, see][and references therein]{King2003}, but their luminosities are always less than $3\times10^{39}\ergs$ \citep[e.g.,][]{Tsygankov1702,Chandra2006}.

The choice of the threshold ULX luminosity at $10^{39}\ergs$ is rather unfortunate because it fails to separate the ``standard'' X-ray binaries (XRBs) from the ``generic'' ULX. Indeed, at least five transient galactic XRBs reach luminosities superior to $10^{39}\ergs$ \citep{Tetarenko1602}. \citet{Middleton1503} suggest that one should take $3 \times 10^{39}\ergs$ as the minimum luminosity defining ULXs whose nature is ``contentious'' (i.e., that could contain either stellar-mass accretors or IMBHs). The idea was to make sure that if the accretor is a black hole (BH), then such a floor value would make sure that the apparent luminosity is super-Eddington. However, this suggestion does not take into account that it is now known that stellar evolution can produce BHs with masses $\gtrsim 30\Msun$ \citep{Belczynski1006, Abbott1602}, and up to $50 \Msun$ under favorable conditions \citep[e.g.][]{Belczynski2012}. Therefore, only sources with luminosities larger than $\sim6\times10^{39}\ergs$, for example, can be expected to be (apparently) super-Eddington if they contain a stellar-mass BH\footnote{The limit may even be higher for He-rich accretion and most massive BHs \citep[e.g.][]{Belczynski2012}}.

Since the isotropic luminosity of an accreting body can be, at best, a few times its Eddington value \citep[see, e.g.][]{Shakura73,Poutanen0705}, observed luminosities $L\gg\Ledd$, where $\Ledd$ is the Eddington luminosity, are only apparently super-Eddington and must be due to radiation beamed toward the observer \citep{King0105,Abramowicz0501}. For example, in the model developed by \citet{King0902} radiation is collimated for $L \gtrsim 3 \Ledd$. One should realise that the beaming of apparently super-Eddington luminosities is unavoidable for neutron star (NS) accretors. In the case of NS ULXs, it has been speculated that magnetar-strength magnetic field ($B > 10^{13}$ G), by increasing the value of the threshold luminosity, would make the emission effectively sub-critical \citep[cf][]{Tong1504,Dallosso1505,Eksi1503,Mushtukov1512}. However, this hypothesis is contradicted by observations in several ULXs of cyclotron resonance scattering feature (CRSF), which corresponds to magnetic fields $B \sim 10^{11}-10^{12}$ G \citep{Brightman1802,Middleton1906, Walton1804}, as well as by the absence of magnetars in binary systems, which is consistently explained by their formation scenarios invoking the destruction of the parent binary system \citep[see][and references therein]{King1905}. The importance of beaming in the context of the ULX origin and observed populations was recently analyzed by \citet{Wiktorowicz1904}.

Since the universal presence of extremely strong magnetic fields in NS ULXs  is very unlikely and is impossible in BH ULXs, the observed luminosities  $\gtrsim3\times10^{39}\ergs$ for NU LXs \citep[we allow for the existence of NSs with masses $\approx2.5\Msun$,][]{Abbott2006}, and larger than $\sim6\times10^{39}\ergs$ in general, cannot be intrinsic and must be beamed, as predicted by \citet{King0105}. This requires high accretion rates\footnote{The relevant accretion rate is that in the external part of the accretion flow where wind mass-loss is negligible.} typically $\dot m\equiv \dot M/\Medd \gtrsim 9$, where $\dot{M}_{\mathrm{Edd}}=2.0 \times 10^{-8} M_{\star}(0.1 / \eta) \Msun/yr$, with $\eta$ being the accretion efficiency and $M_{\star}$ the accretor mass in solar units. Therefore, the accretion rate in bright ULXs ($L_{\rm X}>\sci{6}{39}\ergs$) should be larger than $\sim 1.8 \times 10^{-7} M_{\star}(0.1 / \eta) \Msun/yr$. This is roughly the minimum rate at which accretion discs in (quasi)steady ULXs must be fed by the companion star. Some ULXs are transient \citep[see, e.g.][]{Earnshow2003,Brightman2006}. If this luminosity variability is due to the disk thermal-viscous instability, then the MT rate should be lower than the value mentioned above \citep{Hameury2010}.

The detection of a donor and its orbital motion can shed more light on the MT rates in a ULX and the nature of these objects. However, all ULXs, except potential Galactic ones such as SS433 and the Be-X source Swift J0243.66124, are extragalactic objects and their donors, if not very luminous, remain undetectable for contemporary surveys. Companion stars were dynamically confirmed in the case of two systems: $10$--$20\Msun$ B9Ia star in NGC7793 P13 \citep{Motch1410} and $>3.4\Msun$ WR star in NGC 101 ULX-1 \citep{Liu1311}. Binary population studies suggest that the majority of ULXs are powered by low-mass and low-luminous donors, which is consistent with the lack of detection of any companions in the majority of ULXs. For example, \citet{Wiktorowicz1709} showed that the majority of ULXs will harbour $M<3\Msun$ MS donors. Such stars are only observable in the vicinity of the Sun (even HST is constrained to the Milky Way in their case). Hyper-luminous X-ray sources, which are typically defined as ULXs with apparent luminosities above $10^{41}\ergs$, are still viable candidates to harbour massive donors \citep[$\gtrsim10\Msun$; e.g.][]{Wiktorowicz1509}.

Several ULXs possess donor candidates spatially that are coincident with X-ray sources. Most of them were detected through observations in the near-infrared (NIR) spectral band. In this band, red supergiants (RSG) should easily overcome NIR emission from an accretion disk \citep{Copperwheat0509}. Therefore, these stars were of particular interest in most of the surveys \citep[e.g.][]{Heida1408,Lopez1707,Lopez2009}. Although blue supergiants were, at first, perceived as viable candidates for donors in ULXs because of their prevalence in star forming regions, where most of the ULXs are being found, it was later realised that the optical companions might be outshone by the emission from an irradiated accretion disk \citep[e.g.][]{Sutton1411} unless the separation is very large \citep{Copperwheat0704}. This bias toward high-luminosity NIR companions might be the reason why the majority of the detected companion candidates are RSGs \citep[e.g.][]{Heida1910}. This is further complicated by the fact that we cannot measure distances to ULXs and RSGs directly and the distance to the coinciding galaxy is used as a proxy.

Even if RSG donors form only a small fraction of the intrinsic ULX population, most of these binary companions cannot be Roche-lobe (RL) filling stars. For example, \citet{Wiktorowicz1709} estimated that only $<1\%$ of all ULXs can be accompanied by supergiants if RLOF accretion is concerned exclusively. In binaries containing a compact object and a RSG companion, the mass ratio is typically high ($q=M_{\rm don}/M_{\rm acc}>3$), especially for a NS accretor, which makes the RLOF dynamically unstable \citep[e.g.][]{vandenHeuvel1711} and leads to a common envelope phase \citep[e.g.][]{Ivanova1302}. 

One should stress, however, that $q>3$ may not imply dynamical instability if the stellar envelope is radiative. In this case, which corresponds to the initial phase of Case B binary mass exchange, the MT rate occurs at the thermal (Kelvin--Helmholz) time scale \citep{Kippenhahn6701,Kippenhahn67012} and is equal to
\begin{equation}
\dot{M} \equiv \frac{M_{don,\star}}{t_{\mathrm{KH}}}=3.2 \times 10^{-8} \frac{R_{don,\star}L_{don,\star}}{M_{don,\star}}\,\Msun\,\mathrm{yr}^{-1}, 
\end{equation}
where $M_{\rm don,\star}$, $R_{\rm don,\star}$, and $L_{\rm don,\star}$ are the mass, radius and luminosity of the donor, and index $\star$ designates solar units \citep{Paczynski7101}. This transfer mechanism might power the brightest ULXs \citep{Wiktorowicz1709}. We note that some simulations predict a wider range of binary parameters that give stable mass transfer \citep{Pavlovskii1702}.

Previous theoretical studies mostly neglected a possibility that ULXs can be ``fed'' by stellar winds because, in general, stellar winds are fast and tenuous, so only a small fraction can reach the vicinity of the accretor \citep[see e.g.][]{Pakull0601,Copperwheat0704}. However, RSGs undergo a significant mass loss in stellar wind, which can be slow in the vicinity of the donor, and can therefore, provide a significant wind-fed mass transfer (MT) rate if the separation is small enough or if the wind is focused toward the accretor \citep{elMellah1902}. The presence of an accretion disk in at least two wind-fed XRBs was inferred from spectroscopic observations and modeling (Vela X-1, \citeauthor{Liao2003} \citeyear{Liao2003}; and Cyg X-1, \citeauthor{Zdziarski1410} \citeyear{Zdziarski1410}).

Recently, \citet{Heida1910} suggested that the growing number of ULXs with RSG donor candidates prompts us to reevaluate the significance of wind-driven MT in the context of ULXs. However, to date, no general population studies of wind-fed ULXs have been done, even though at least one ULX was detected with a donor that is not filling its RL  \citep[e.g. M 101 ULX-1,][]{Liu1311}. The aim of this article is to tackle this problem. In particular, we will answer the following question: how important is the wind-fed accretion for ULX sources? Specifically, we perform a statistical analysis of binary evolution models in the context of recent observations in order to derive the number and properties of wind-fed ULXs. We focus on NS ULXs with RSG donors and progenitors of gravitational wave sources, such as GW170817.

\section{Methods}

In this section we will describe our adopted models (modes) of wind accretion. The inclusion of wind accretion in XRB models is the most significant difference in respect to the previous population studies of ULXs. Bondi--Hoyle--Lyttleton (BHL) accretion is a standard scheme of wind mass accretion in \startrack\ \citep{Belczynski0801,Belczynski2004}. Additionally, we adapt the wind RLOF (WRLOF) scheme, which was already used by \citet{Ilkiewicz1906} to analyze the relation between SNIa and wide symbiotic stars. 

The main source of uncertainty for accretion models is actually the origin of the wind mass--loss from the donor star. The standard model for stellar winds as implemented in \startrack\ is described in \citet{Belczynski1005}. In short, we use the wind mass-loss prescriptions for OB stars based on observational data \citep{Vink0104,Vink0810}. For Wolf-Rayet wind, the code accounts for clumping \citep{Hamann9807} and metallicity dependence \citep{Vink0511}. Wind mass-loss rates for luminous blue variable stars are calculated from $\sci{1.5}{-4}(M/\Msun)\msy$ \citep{Vink0210,Belczynski1005}. For low-mass stars, we follow the formalism of \citet{Hurley0007}. This prescription is a minimal realistic one as far as the scarcity of the available observational data is concerned, but we note that the actual model realised in Nature may diverge from what we employed. This should be investigated by future observational campaigns.

\subsection{Bondi--Hoyle--Lyttleton accretion}

In the BHL scheme \citep[for a recent review see \citeauthor{Edgar0409} \citeyear{Edgar0409}]{Hoyle3901,Bondi5201}, we assume that the accretor is engulfed by a steady and uniform (at least in the vicinity of the accretor) supersonic wind from a companion star with $v_{\rm wind}\gg v_{\rm orb}$, where $v_{\rm wind}$ is the wind velocity relative to the accretor and $v_{\rm orb}$ is the orbital velocity of the accretor relative to the mass losing star. We calculate the fraction of the wind mass-loss accreted by the compact body as \citep[for derivation see][Equation~39]{Belczynski0801}
\begin{flalign}\label{eq:dotMwind}
    \nonumber \beta_{\rm acc,BHL}=&\frac{\dot{M}_{\rm acc,BHL}}{\dot{M}_{\rm don,wind}} = \frac{1}{\sqrt{1-e^2}}\times\\
                     &\left(\frac{G M_{\rm acc}}{v^2_{\rm wind}}\right)^2
                     \frac{\alpha_{\rm wind}}{2 a^2} 
                     \left(1+\left(\frac{v_{\rm orb}}{v_{\rm wind}}\right)^2\right)^{-\frac{3}{2}},   
\end{flalign}
\noindent where $\dot{M}_{\rm acc,BHL}$ is the orbit-averaged wind accretion rate, $e$ is the eccentricity, $G$ is the gravitational constant, $M_{\rm acc}$ is the accretor's mass, $a$ is the binary separation, and $\dot{M}_{\rm don,wind}$ is the mass-loss rate in wind from the donor. $\alpha_{\rm wind}$ is a numerical factor between $1$ and $2$ \citep{Boffin8810,Shima8511}. Following \citet{Belczynski0801}, we assume a conservative value of $\alpha_{\rm wind}=1.5$. We impose a limit of $\dot{M}_{\rm acc,BHL}\leq0.8\times\dot{M}_{\rm don,wind}$ to avoid artificially high values resulting from orbital averaging. A formula similar to Equation \ref{eq:dotMwind} was first proposed by \citep{Bondi5201}, but later \citet{Shima8511} showed that an additional factor of $\sim2$ makes the Bondi's formalism fit their hydrodynamical simulations and agree with the earlier simplified prescription of \citet{Hoyle3901}. 

Here we make an optimistic assumption that all of the mass that is captured into the wake is consequently accreted, although only a fraction of this mass is expected to be accreted in a real situation \citep[e.g.][]{Edgar0409}. We also do not include any effects of orbital motion on the accretion flow, which could additionally decrease the accretion rate  \citep[e.g.][]{Theuns9606}. Consequently, the mass-accretion rate calculated through Equation~\ref{eq:dotMwind} should be perceived as an approximate upper limit and our results should be seen as optimistic.

\subsection{Wind RLOF}\label{sec:wrlof}

When the stellar wind is slow inside the donor's RL, the assumption of wind isotropy may not be realistic because the accretor can gravitationally focus the wind toward the L1 point, thus imitating RLOF-like mode of accretion (including formation of an accretion disk). In such a case, one speaks about WRLOF. \citet{Mohamed0709} showed that it can provide much higher rates than BHL accretion in favourable circumstances. According to their simulations, which are limited to few special cases, WRLOF MT rate can be even 100 times higher than for BHL. The WRLOF has been already used to explain the observations of symbiotic stars \citep[e.g.,][]{Mohamed0709,Ilkiewicz1906}, carbon-enhanced metal-poor stars \citep{Abate1304}. In particular, \citet{elMellah1902} showed that in favourable conditions wind-powered XRBs with supergiant donors can appear as ULXs in favourable conditions.

The MT may resemble the standard RLOF in detached binaries containing stars that have slow, dust-driven wind  \citep{Mohamed0709}. The slow moving material fills the RL of the donor (stellar wind source), before being significantly accelerated by the radiation pressure and escaping preferentially through L1 point. Therefore, a significant fraction of the wind may be collimated toward the orbital plane and the accretor. 

The formalism for WRLOF used in the present study, which we describe below, was first developed by \citet{Abate1304}. Motivated by numerical simulations \citep[e.g.][]{Mohamed10}, we assume that the collimation of stellar wind toward the orbital plane and the accretor is significant when the radius where the stellar wind accelerated by the radiation pressure reaches the escape velocity ($R_{\rm d}$) is comparable to or larger than the RL radius. Assuming $R_{\rm d}\gg R_{\rm don}$, the former can be approximated as \citep{Hofner07}:
\begin{equation}
    R_{\rm d}=\frac{1}{2}R_{\rm don}\left(\frac{T_{\rm eff}}{T_{\rm cond}}\right)^{\frac{4p}{2}},
\end{equation}
\noindent where $R_{\rm don}$ is the donor's radius, $T_{\rm eff}$ the donor's effective temperature, $T_{\rm cond}$ is the condensation temperature of the dust, and $p$ is a parameter. Although $T_{\rm cond}$ and $p$ depend on the chemical composition, we assume the values of $T_{\rm cond}=1500$ K and $p=1$, which are consistent with properties of carbon-rich dust \citep[e.g.][]{Hofner07}.

The stellar-wind fraction reaching the accretor is calculated as \citep[see Equation~9 therein]{Abate1304}:
\begin{equation}
    \beta_{\rm acc,WRLOF}=\min\left\{\left(\frac{q}{0.6}\right)^2\left[c_1x^2+c_2x+c_3\right],\beta_{\rm acc,max}\right\},
\end{equation}
\noindent where $x=R_{\rm d}/R_{\rm RL,don}$, $R_{\rm RL,don}$ is the RL radius of the donor, $c_1=-0.284$, $c_2=0.918$, and $c_3=-0.234$. $c_1x^2+c_2x+c_3$ is a fit to the results of \citet{Mohamed10} made for $q=0.6$, $q$ being the mass ratio (donor mass over accretor mass). $(q/0.6)^2$ is the scaling proposed by \citet{Abate1304} to account for the influence of the mass ratio on the MT rate. $\beta_{\rm acc,max}=0.5$ is the highest value of $\beta_{\rm acc}$ obtained by \citet{Mohamed10}, which we use as the limit for all values of $q$. If the MT rate calculated for the WRLOF is higher than the BHL rate, then we assume that MT occurs through WRLOF mode. In the opposite case, we assume that it occurs thought BHL mode.

We are aware that the simulation of \citet{Mohamed10} and calculations by \citet{Abate1304} were performed for a narrow region of the parameter space (viz. specific values of accretor and donor masses). Nonetheless, we extended the applicability of their results to suit the needs of population synthesis calculations. In our reference model (WM8), we allow for the WRLOF MT mode only from donors less massive than $8\Msun$, which is just the maximal mass to which simulations were performed in \citet{Mohamed10}. For comparison we provide also the results for two other models. In the WM0 model the WRLOF MT is not possible and all wind-fed systems transfer mass using BHL mode. Meanwhile, in the model WM+, WRLOF MT is allowed for all donors irrespective of their mass. These two variations may be seen as extreme cases as far as the importance of WRLOF is concerned.

\subsection{Simulations}

To obtain statistically significant comparison of observational data on ULX sources with models of wind-fed accretion, we utilized the population synthesis code \startrack\ \citep{Belczynski0801,Belczynski2004}. In a previous paper \citep{Wiktorowicz1904}, we have analyzed the observed population of ULXs but focused on the RLOF systems only. In the current work, ULXs in which accretion occurs via RLOF serve as a reference point. The beaming (i.e., anisotropic emission of radiation) can be present in ULXs in general \citep[e.g.][]{Wiktorowicz1904} and specifically in PULXs \citep{King2003}. Therefore, we include these effects in our analysis of wind-fed ULXs as was previously done for the population of RLOF ULXs in \citet{Wiktorowicz1904}.

To compare the theoretical binary evolution models and observational data, we calculate the predicted number of RLOF-fed and wind-fed ULXs, for a fiducial galaxy that resembles the Milky Way. Specifically, we assumed the total stellar mass to be $M_{\rm tot}=\sci{6.08}{10}\Msun$ \citep{Licquia1506}, a uniform metallicity distribution ($Z=\Zsun=0.02$) and constant star formation rate (SFR) throughout the last $10\Gyr$ equal to $6.08\msy$. We also present results for burst-like star formation episodes ($\Delta t=100\Myr$; SFR=$608\msy$) for comparison at two moments: $100\Myr$ and $2\Gyr$ after the commencement of star formation (i.e. right after and $1.9\Gyr$ after  he star formation ceases), which correspond to young and old stellar populations, respectively. Due to the inaccuracies of evolutionary models and sparse observational data, a more-detailed Galactic model is not necessary in the context of this work.

\section{Results}

\subsection{Observable sample}

\begin{deluxetable*}{l|ccc|ccc|ccc}
    \tablewidth{\textwidth}
    \tablecaption{Number of ULXs for a Fiducial Galaxy}

    \tablehead{Model & \multicolumn{3}{c}{Constant SF} & \multicolumn{3}{c}{Burst SF ($100\Myr$ ago)} & \multicolumn{3}{c}{Burst SF ($2\Gyr$ ago)}\\
    & RLOF & WIND & $f_{\rm WRLOF}$ & RLOF & WIND & $f_{\rm WRLOF}$ & RLOF & WIND & $f_{\rm WRLOF}$}
    \startdata
    
\multicolumn{10}{c}{All Companions}\\
WM0          & \sci{2.4}{1} & \sci{2.3}{1} & -- & \sci{6.8}{2} & \sci{2.1}{3} & -- & \sci{2.2}{1} & \sci{3.7}{-2} & -- \\
WM8          & \sci{2.4}{1} & \sci{3.7}{1} & 42.50\% & \sci{6.8}{2} & \sci{2.9}{3} & 33.63\% & \sci{2.2}{1} & \sci{1.7}{-2} & 100.00\% \\
WM+        & \sci{2.4}{1} & \sci{1.9}{2} & 97.24\% & \sci{6.8}{2} & \sci{1.8}{4} & 97.96\% & \sci{2.2}{1} & \sci{1.7}{-2} & 100.00\% \\
\multicolumn{10}{c}{Supergiant Companions}\\
WM0          & \sci{5.3}{-2} & \sci{2.0}{1} & -- & 5.2 & \sci{1.8}{3} & -- & -- & -- & -- \\
WM8          & \sci{5.3}{-2} & \sci{3.3}{1} & 44.45\% & 5.2 & \sci{2.7}{3} & 36.08\% & -- & -- & -- \\
WM+        & \sci{5.3}{-2} & \sci{1.8}{2} & 98.60\% & 5.2 & \sci{1.8}{4} & 99.33\% & -- & -- & -- \\
\multicolumn{10}{c}{RSG companions}\\
WM0          & \sci{1.6}{-5} & \sci{2.0}{1} & -- & \sci{1.6}{-3} & \sci{1.8}{3} & -- & -- & -- & -- \\
WM8          & \sci{1.6}{-5} & \sci{3.2}{1} & 43.51\% & \sci{1.6}{-3} & \sci{2.6}{3} & 35.57\% & -- & -- & -- \\
WM+        & \sci{1.6}{-5} & \sci{1.8}{2} & 98.56\% & \sci{1.6}{-3} & \sci{1.7}{4} & 99.32\% & -- & -- & -- \\

    \enddata
    \tablecomments{Predicted numbers of ULXs powered by Roche lobe overflow (RLOF) and wind accretion (WIND) in a fiducial galaxy (see text for details). $f_{\rm WRLOF}$ gives a fraction of wind-fed ULXs in which MT occurs through wind RLOF mode (Section~\ref{sec:wrlof}). Models presented are: WM0 in which only BHL accretion mode is possible, WM8 in which WRLOF accretion mode possible, but limited to donors lighter than $8\Msun$, and WM+ in which WRLOF accretion mode possible for donors of all masses. Additionally, supergiant and RSG donors are shown separately. Supergiant companions are defined as post-MS stars with luminosity $L>15000\Lsun$ and surface gravity $\log g<2$, whereas RSGs, a subpopulation of supergiants, additionally have effective surface temperatures lower than $4$kK. Results for three reference SFHs are presented: constant through $10\Gyr$ and burst-like with a duration of $100\Myr$ which started $100\Myr$ or $2\Gyr$ ago. Zeros are marked with '--' for clarity.}
    \label{tab:results}
\end{deluxetable*}

Table~\ref{tab:results} presents the abundance of ULXs (both RLOF- and wind-fed) in our results. Although these numbers are rather optimistic due to favorable assumptions, particularly the  high recent SFR used in the simulations and the omission of interstellar absorption, they show that the inclusion of wind-fed sources in ULX populations studies is necessary. Wind-fed ULXs may not only represent a significant fraction of the ULX population (e.g. $49$--$89\%$ for constant star formation history) but, in favorable conditions (e.g. very young stellar environments), may also be its dominant component. We note that these results may change significantly if the actual wind-emission model from massive stars differs from that used in the current study.

\begin{figure}
    \centering
    \includegraphics[width=\columnwidth]{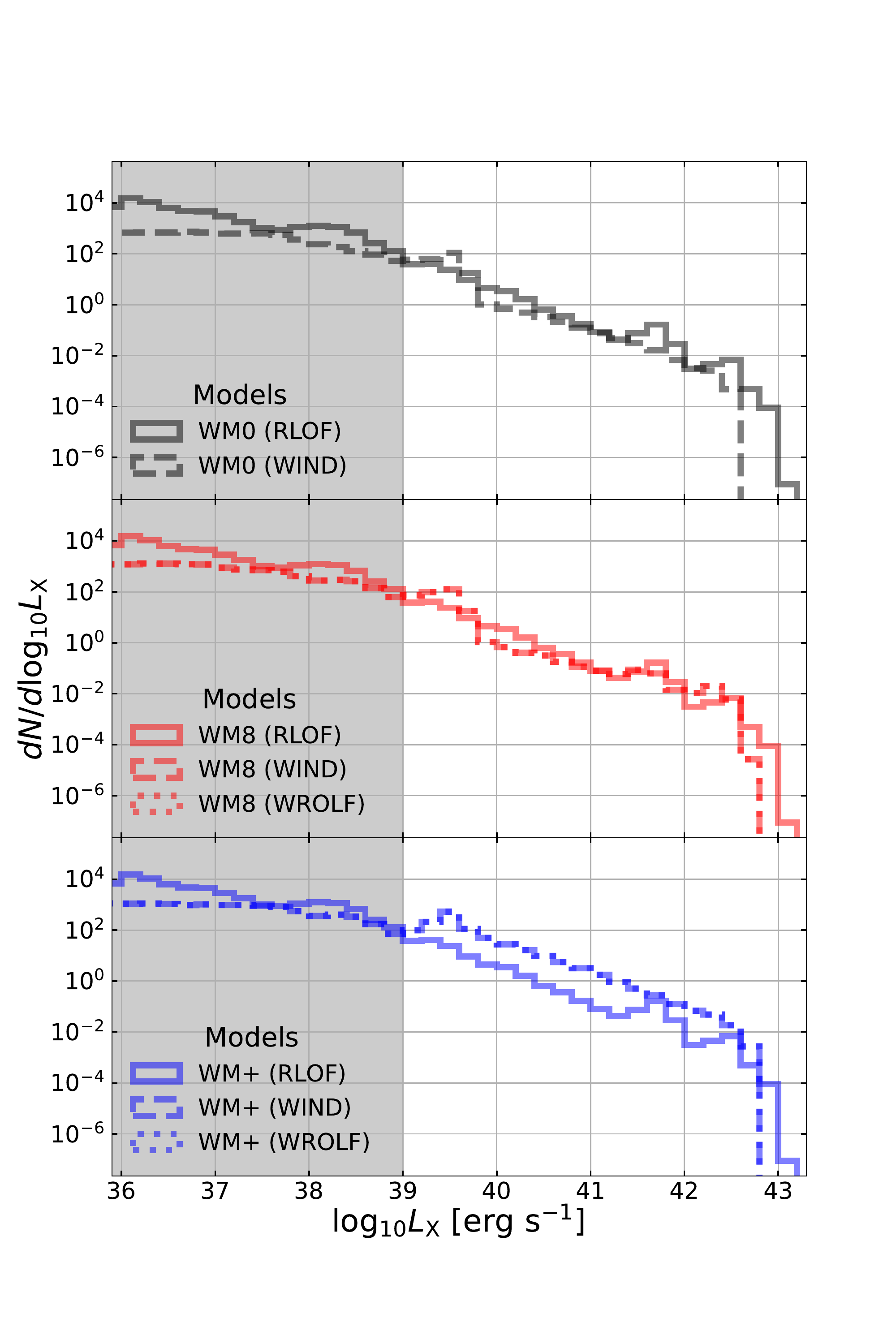}
    \caption{Distributions of X-ray luminosity ($L_{\rm X}$) of ULXs for the various models tested in this study (see text for details). The shaded area presents the extension for regular XRBs ($L_{\rm X}<10^{39}\ergs$), but is not a part of this study. Designations in parenthesis mark the ULX subpopulations: RLOF--powered by the RLOF accretion mode; WIND--powered by wind-fed accretion (both WRLOF and BHL modes); WRLOF--powered by WRLOF accretion mode only. }
    \label{fig:Lxdist}
\end{figure}

The population of ULXs is dominated by sources with low luminosities ($\lesssim3\times10^{39}\ergs$) both for RLOF and wind-powered sources (Figure~\ref{fig:Lxdist}). These are mainly BH ULXs emitting isotropically and mildly-beamed (beaming factor\footnote{The beaming factor $b$ is the ratio of the solid angle of emission and the whole sphere. In this study we assume $b=\min(1, 73/(\dot{M}/\dot{M}_{\rm Edd})^2)$, where $\dot{M}$ and $\dot{M}_{\rm Edd}$ are the actual and the Eddington accretion rates, respectively \citep{King0902,Wiktorowicz1904}.} $b\gtrsim0.2$) NS ULXs. In the case of wind-fed ULXs, both modes (viz. BHL and WRLOF) are possible. As far as the most luminous sources are concerned, the highest luminosities (up to $\sim10^{43}\ergs$) are obtained for binaries consisting of a NS with a Helium rich donor or a BH with a HG donor \citep[see also][]{Wiktorowicz1509}.  In the case of wind-fed sources, this highly luminous group is composed of ULXs with BH accretors transferring mass through the WRLOF mode exclusively. In all cases, we assume that the beaming is saturated at $b\approx b_{\rm min}=3.2\times10^{-3}$ \citep{Lasota1603,Wiktorowicz1709}. We note, that no ULXs have been observed with luminosities much above $\sim10^{42}\ergs$ and some ULXs with luminosities above $\sim10^{41}\ergs$ might actually contain intermediate-mass BHs \citep[e.g.][]{Greene1911}, which are not included in our study. Nonetheless, although these luminous stellar-mass ULXs are rare (we predict $\sim10^{-7}$ per Milky Way equivalent galaxy), they may exist undetected in distant galaxies. \citet{King0902} suggested that such highly beamed and apparently luminous sources may mimic distant AGNs acting as ``pseudoblazars.''

\begin{figure*}
    \centering
    \includegraphics[width=\textwidth]{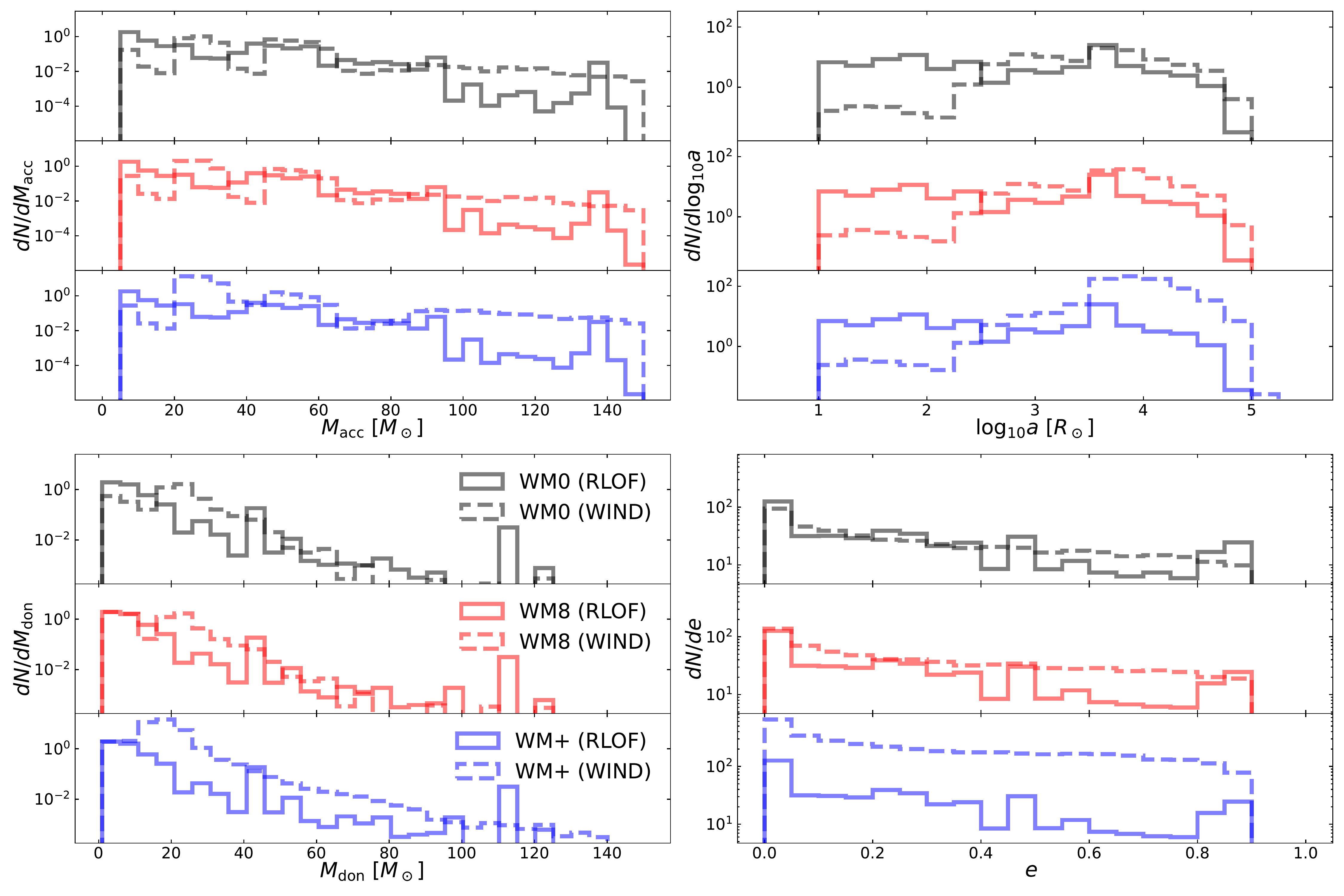}
    \caption{Initial distributions of parameters for ULX progenitors with division on different tested models and accretion modes (see caption of Figure~\ref{fig:Lxdist}). Parameters are: $M_{\rm acc}$--accretor mass; $M_{\rm don}$--donor mass; $a$--separation; $e$--eccentricity.}
    \label{fig:ZAMSdists}
\end{figure*}

The wind-fed ULXs can reach similar luminosities as RLOF systems, despite the large fraction of mass being lost from the system \citep[e.g.][obtained that no more than $50\%$ of wind is accreted]{Mohamed10}. This is possible because the wind-fed systems do not suffer from the dynamical instability when the mass ratio is high ($\gtrsim3$), even when their envelopes are convective \citep[][and references therein]{vandenHeuvel1711}, and they can have very massive donors. Consequently, the progenitors of wind-fed ULXs are typically more massive on the ZAMS and have larger separations to accommodate expanding stars (Figure~\ref{fig:ZAMSdists}).

\begin{figure}
    \centering
    \includegraphics[width=\columnwidth]{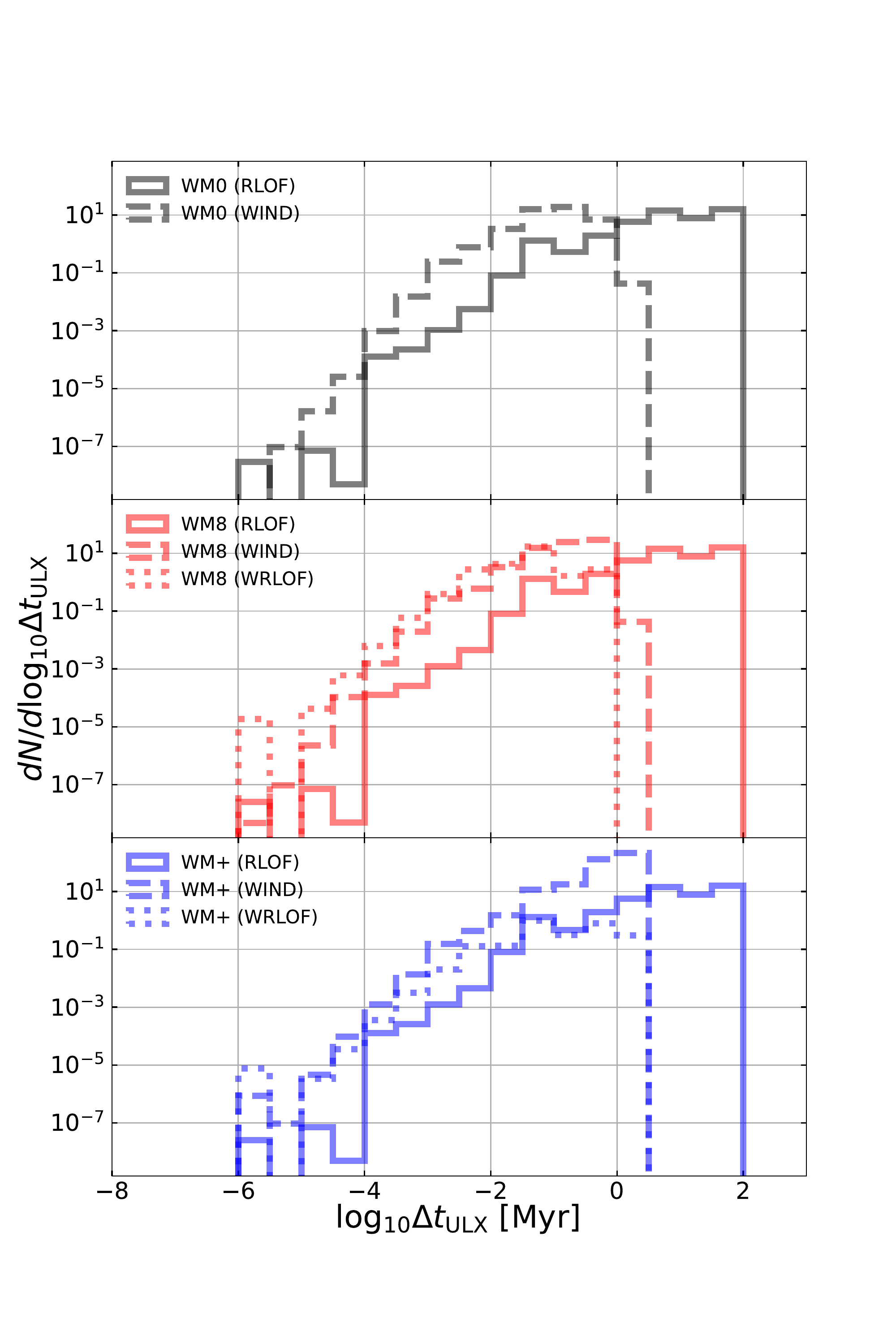}
    \caption{Distribution of the ULX phase total duration times ($\Delta t_{\rm ULX}$) with a division on different models and accretion modes (see the caption of Figure~\ref{fig:Lxdist}). The durations are combined when a system has more than one ULX phase during its evolution.} 
    \label{fig:dtdist}
\end{figure}

The RLOF powered ULXs typically have much longer duration of the emission phase ($>5\Myr$) than wind-fed systems ($\sim1\Myr$ for WRLOF and $\sim0.1\Myr$ for BHL; see Figure~\ref{fig:dtdist}). The donors in wind-fed systems are typically massive stars (Figure~\ref{fig:donordists}) in short-lived supergiant phase (e.g. $\sim1.5\Myr$ for $5\Msun$ asymptotic giant branch star (AGB) or $\sim0.1\Myr$ for $10\Msun$ AGB star) when their wind is relatively slow and dense. Meanwhile, RLOF accretion may give high MT rates even for low-mass companions at the end of MS phase \citep[e.g.][]{Wiktorowicz1709}. Low-mass stars have significantly longer life spans than their heavier cousins. Therefore, they potentially have longer MT phases and less often enter dynamical-time MT when their RL is filled. We note that it is only possible for the donor in a RLOF-powered ULX to fill its RL immediately after the compact object formation as a result of favorably directed natal kick in rare situations \citep[see e.g.][]{Maxted2010}, so that its MT phase can be the longest possible. In a typical case, the RL is filled only after the star expands as a result of nuclear evolution, which is typically shortly ($\lesssim100\Myr$) before TAMS.   

RLOF ULXs are typically associated with low- and medium-mass donors ($\lesssim10\Msun$), whereas wind-fed ones typically have massive donors ($\gtrsim10\Msun$; Figure~\ref{fig:donordists}). Although low-mass stars are more abundant among primaries due to the shape of the initial mass function ($\Gamma<0$), flat initial mass ratio distribution makes massive companions more prevalent among NS/BH progenitors \citep[Figure~\ref{fig:ZAMSdists}; see also][]{Wiktorowicz1911}. This situation favors wind-fed ULXs. The higher the mass of a star, the stronger is the mass loss in stellar wind for a particular evolutionary phase. This makes a binary with compact object progenitor and massive secondary a perfect predecessor of a wind-fed ULX. Moreover, when a massive star fills its RL after the primary forms a compact object, the mass ratio may be too high to provide a stable RLOF MT and instead a CE occurs and this bars the formation of a ULX. We note that ULXs powered by RLOF MT have been discussed thoroughly in previous studies \citep[e.g.][]{Wiktorowicz1709}.

\subsection{Typical evolution} \label{sec:er}

Wind-fed ULXs are expected to comprise a large fraction (up to $\sim96\%$) of ULXs in star-forming environments, but practically disappear ($<1\%$) in old environments (Table~\ref{tab:results}). These systems may contain either a BH or a NS accretor, and they may be powered by WRLOF or BHL wind accretion mode. However, NS accreting through WRLOF  mode dominate the population in older environments ($\gtrsim1\Gyr$ after the end of star-formation). In this section, we compare various evolutionary routes leading to the formation of wind-fed ULXs and their posterior evolution.

\begin{figure}
    \centering
    NS ULX WRLOF accretion mode\\
    \includegraphics[width=\columnwidth]{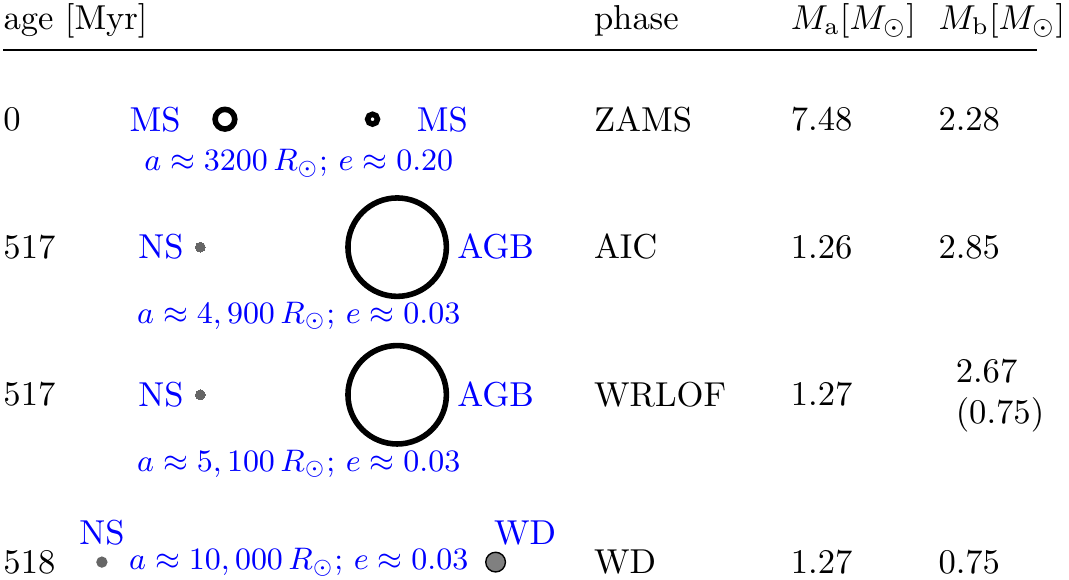}\\
    \vspace{0.5cm}
    NS ULX BHL accretion mode\\
    \includegraphics[width=\columnwidth]{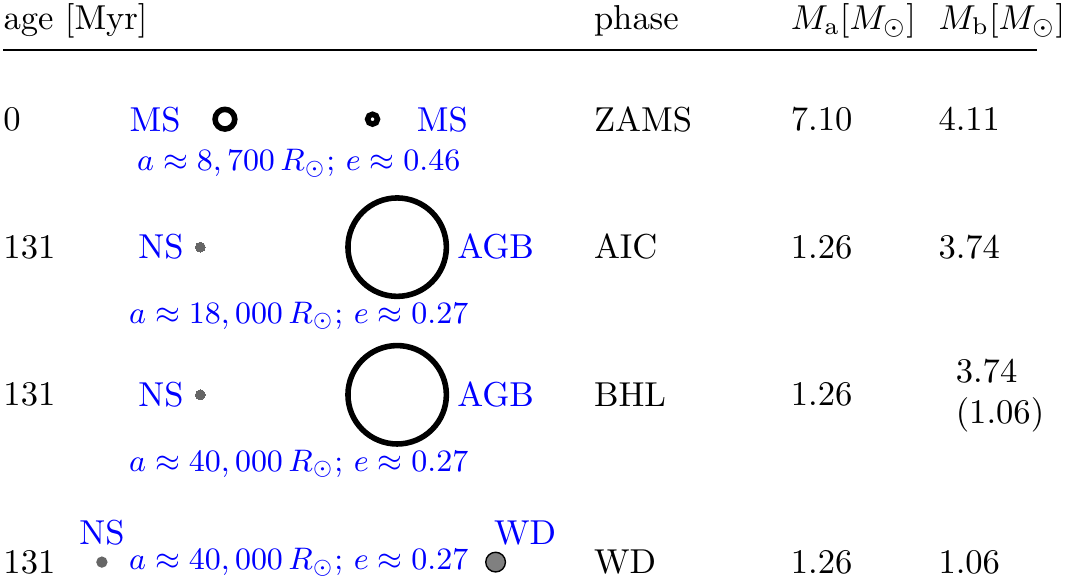}
    \caption{Symbolical representation of a typical system evolution leading to the formation of wind-fed NS ULXs. Upper/Lower plots show routes with WRLOF/BHL accretion mode. The columns represent the age of the system, scheme of the system (not to scale), evolutionary phase, and masses of the primary and secondary. The binary evolution phases are as follows: ZAMS--zero age main sequence; AIC--accretion induced collapse; WRLOF--phase of wind accretion through WRLOF mode; BHL--phase of wind accretion through Bondi--Hoyle--Lyttletone mode; WD--white dwarf formation. Additionally, the schemes present separations ($a$), eccentricities ($e$), and stellar evolutionary phases: MS--main sequence; AGB--asymptotic giant branch; NS--neutron star; WD--white dwarf.}
    \label{fig:ts_NS}
\end{figure}

The most common wind-fed ULXs are those with NS accretors and MT occurring through the WRLOF mode. Donors in these systems are typically low-mass ($0.7$--$2.6\Msun$) AGB stars. In a typical case (Figure~\ref{fig:ts_NS}, upper plot), the binary on ZAMS composes of a $7.48\Msun$ primary and a $2.28\Msun$ secondary on a medium-size orbit of $\sim3200\Rsun$ with moderate eccentricity ($e\approx0.2$). The primary evolves and becomes an AGB star at the age of $48\Myr$. The stellar wind leads to a significant mass loss, part of which is being captured by the secondary. Finally, the primary becomes a $1.36\Msun$ ONe white dwarf (WD), whereas the secondary grows to $\sim3\Msun$ while still being on the MS. After $460\Myr$ the secondary evolves off the MS and becomes an AGB star, which results in a significant rise in the rate of mass loss in stellar wind. The WD captures a fraction of the wind, which allows it to grow to $1.38\Msun$, when the accretion induced collapse transforms it into a $1.26\Msun$ NS. The secondary remains an AGB star for the next $0.7\Myr$, while the mass loss increase and in a short time reaches $\sim10^{-6}\msy$ out of which $37\%$ is captured into the primary disk. This marks the beginning of the ULX phase that lasts for $100\kyr$ and the luminosity reaches $\sim\sci{4}{41}\ergs$. Afterwards, the secondary becomes a $0.75\Msun$ CO WD.

The progenitors of NS ULXs powered through the BHL mode require heavier companions ($1.2$--$2.4\Msun$) and larger initial separations. In a typical case (Figure~\ref{fig:ts_NS}, lower plot), the system initially consist of a $7.1\Msun$ primary and $4.1\Msun$ secondary on a wide eccentric orbit ($a\approx8700\Rsun$; $e\approx0.46$). The primary, while ascending the AGB, loses most of its mass and at the age of $54\Myr$ becomes an ONe WD. As a result of mass accretion, the secondary grows to $\sim5.5\Msun$ and after $80\Myr$ becomes an AGB star. In one \Myr\ it loses nearly $2\Msun$ of its mass, due to strong stellar wind, a fraction of which is captured by the WD, which increases its mass to $1.38\Msun$ when the electron-capture SN results in the  formation of a NS. In such an accretion induced collapse, the natal kick is expected to be low, which allows the wide binary with a separation of $18,000\Rsun$ and moderate eccentricity of $e\approx0.27$ to survive. Such a large separation makes WRLOF ineffective and the MT occurs through the BHL mode. Although the fraction of the wind being accreted is small ($\sim2\%$), the strong stellar wind ($\sci{6.5}{-5}\msy$) makes the effective MT quite high ($\sci{1.5}{-6}\msy$), comparable to that in typical NS ULX powered through the WRLOF mode (see paragraph above). The luminosity reaches $\sci{3.5}{40}$ for a brief moment, but within $3\kyr$ the system dims and becomes a regular XRB. After $100\kyr$, the secondary becomes a WD with a mass of $\sim1.06\Msun$.

We note that AGB stars with mass-loss rates as in the above examples are observed in the Large Magellanic Cloud \citep[e.g.][]{Hofner1801}. Therefore, the accretion efficiencies in WRLOF mode that are necessary to power a ULX for the observed mass-loss rates are within the model range (i.e. $\leq0.5$). Concerning the BHL accretion mode, we may estimate the accretion radius as $R_{\rm acc}=2GM_{\rm acc}/v_{\inf}^2$, where $G$ is the gravitational constant and $v_{\inf}$ is the terminal wind velocity \citep[e.g.][]{Edgar0409}. The fraction of accreted wind may be further estimated as the fraction of the wind that falls within the accretion radius: $f_{\rm acc}=0.5\sin R_{\rm acc}/a\approx0.095\,(M/M_\odot)(v_{\inf} /10\kms)^{-2}(a/20,000\Rsun)^{-1}$, where $v_{\inf}\approx10\kms$ is the typical value for AGB stars \citep[e.g.][]{Hofner1801}. Although these estimations are very rough (see Section~\ref{sec:accretion}), they show that the MT rates in the typical routes are feasible.

\begin{figure}
    \centering
    BH ULX WRLOF accretion mode\\
    \includegraphics[width=\columnwidth]{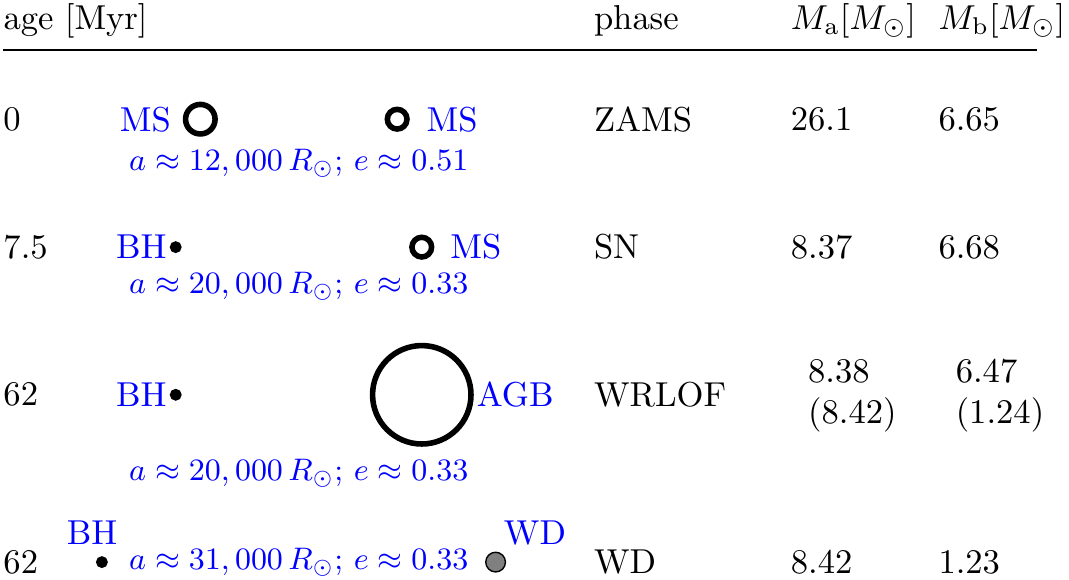}\\
    \vspace{0.5cm}
    BH ULX BHL accretion mode\\
    \includegraphics[width=\columnwidth]{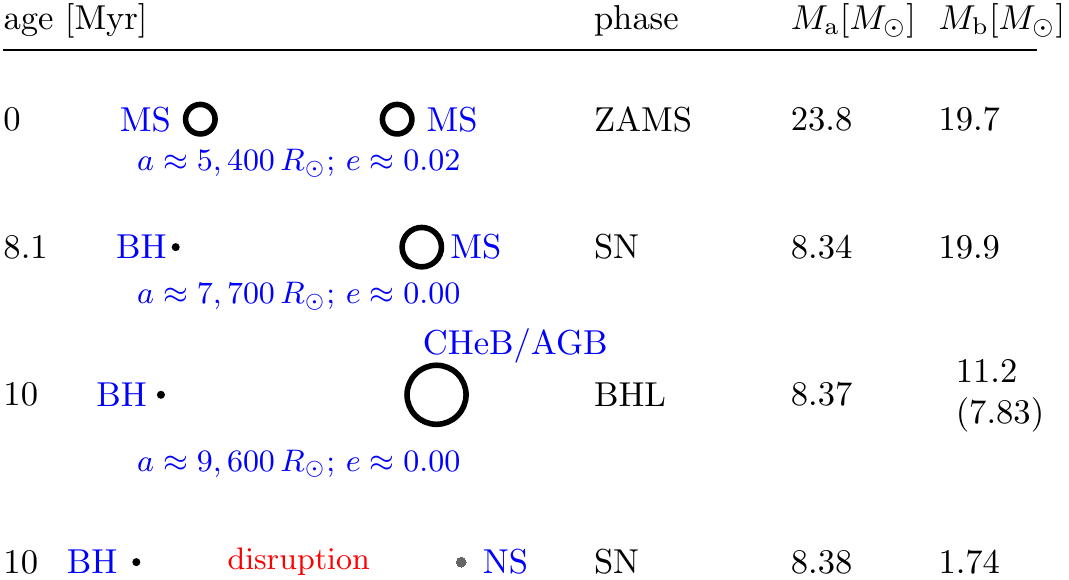}
    \caption{Same as Figure~\ref{fig:ts_NS}, but for ULXs with BH accretors. Additional abbreviations (not present in Figure~\ref{fig:ts_NS}) for binary evolution include: SN--supernova (i.e. formation of the compact object); whereas for stellar evolution: BH--black hole; CHeB--core Helium burning.}  
    \label{fig:ts_BH}
\end{figure}

The companions in wind-powered BH ULXs are (in general) more massive than in NS ULXs. This results from the fact that heavier primaries (BH progenitors) are typically accompanied by heavier secondaries ($\gtrsim10\Msun$) on the ZAMS \citep[e.g.][]{Wiktorowicz1911}. For BH ULXs, in which the MT occurs through WRLOF mode, the typical companion mass is $5$--$7\Msun$, whereas a BH is typically $\sim8$--$9\Msun$. In a typical evolution leading to the formation of such a system (Figure~\ref{fig:ts_BH}, upper plot), the primary on ZAMS is $26.1\Msun$, whereas the secondary is $6.65\Msun$, and the separation is $12,000\Rsun$ with significant eccentricity $e\approx0.51$. The primary evolves more quickly and after $6.7\Myr$ becomes a CHeB star. A high wind mass-loss starts that decreases the star's mass to $9.2\Msun$ in just $700\kyr$. A fraction of this mass is accreted by the secondary, which allows it to retain its initial mass of $\sim6.7\Msun$ despite loses in stellar wind. Shortly after, the primary forms a $8.37\Msun$ BH. Meanwhile, the separation increases to $20,000\Rsun$ and the eccentricity changes to $0.33$. The secondary needs more than $50\Myr$ to evolve off the MS and become an AGB star, when its wind mass loss grows significantly to $\sci{4.6}{-7}$. The wind is still slow inside the RL. Therefore, a large fraction ($\sim50\%$) is gravitationally collimated toward the L1 point and feeds the accretion disk of the primary. For $500\kyr$ the MT through WRLOF mode is high enough to power a ULX with the peak luminosity reaching $\sim10^{42}\ergs$. Afterwards, the secondary becomes a CO WD with a mass of $1.23\Msun$.

The evolution leading to the formation of BH ULXs powered by BHL accretion (Figure~\ref{fig:ts_BH}, lower plot) differs significantly from the cases described above. The companions are significantly more massive, $9$--$12\Msun$, thus the WRLOF can operate only in WM+ model, which allows for wind collimation from donors with masses above $8\Msun$. On the ZAMS, the binary consists of a $23.8\Msun$ primary and a $19.7\Msun$ secondary on an orbit of $5400\Rsun$, which is nearly circular ($e\approx0.02$). The primary evolves and in $7.3\Myr$ becomes a CHeB star, which commences a significant mass loss in stellar wind. Within $800\kyr$ the star loses more than half of its mass in stellar wind and becomes a $8.34\Msun$ BHs. The secondary evolves and at the age of $\sim10\Myr$, as a CHeB star, provides a MT rate of $\sci{1.2}{-5}\msy$ out of which $\sim2\%$ reaches the vicinity of the accretor. It is high enough to power an isotropic ULX ($L_{\rm X}\approx(2$--$3)\times10^{39}\ergs$) for $\sim100\kyr$. Meanwhile, the secondary evolves into an AGB star and the luminosity of the ULX increases to reach $\sim10^{42}\ergs$. After that time, the secondary, having a mass of $7.8\Msun$, undergoes a SN and becomes a NS. However, the NS's natal kick on such a wide orbit ($a\approx13,000\Rsun$) results in a binary disruption. We note that the accretion mode in this scenario is forced by the limit on the maximal mass of donor as $8\Msun$ in WRLOF mode for the reference mode. In WM+ model, this ULXs will transfer mass through WRLOF mode but the peak luminosity will be similar.

\begin{figure}
    \centering
    BH ULX RLOF+BHL accretion mode\\
    \includegraphics[width=\columnwidth]{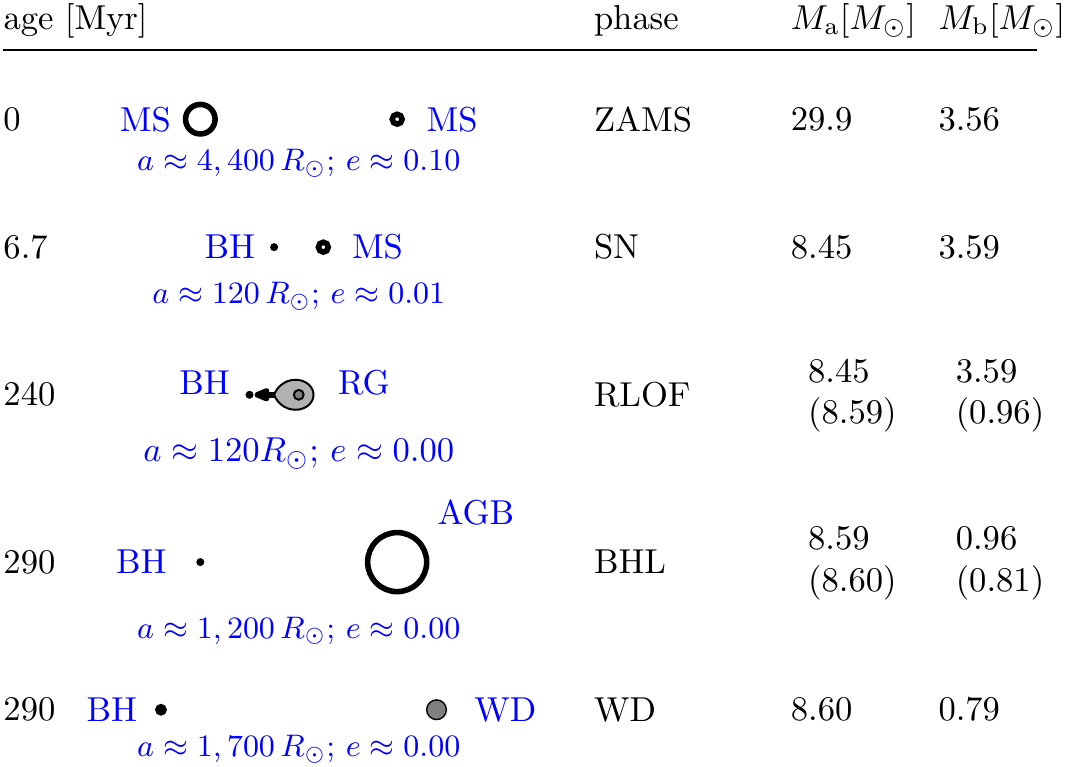}
    \caption{Same as Figure~\ref{fig:ts_NS}, but for a ULX that is powered first by RLOF accretion and later by wind accretion. Additional abbreviations (not present in Figures~\ref{fig:ts_NS} and \ref{fig:ts_BH}) for the binary evolution are: RLOF--Roche-lobe overflow; and for stellar evolution: RG--red giant.}  
    \label{fig:ts_RLOFWIND}
\end{figure}

A few ULXs evolve through both the RLOF powered phase and wind powered phases. This situation occurs for $4\%$ of RLOF ULXs and $5\%$ of wind-powered ULXs ($1\%/8\%$ in the case of WM0/+ models, respectively) predicted to be observed currently in the modeled galaxy. In a typical case (Figure~\ref{fig:ts_RLOFWIND}, such a binary on the ZAMS is composed of a $29.9\Msun$ primary and $3.56\Msun$ secondary, whereas the separation and eccentricity are $4400\Rsun$ and $0.10$, respectively. The primary evolves more quickly and becomes a CHeB star. The mass loss in the stellar wind removes most of the hydrogen envelope, so when the star fills its RL at the age of $6.5\Myr$ the companion easily rejects the CE and the separation shrinks to $100\Rsun$. The primary evolves as a helium star for $200\kyr$ and forms a $8.45\Msun$ BH. The orbit now is close enough for the secondary expanding as a RG to fill its RL and start a RLOF, which is strong enough to power a ULX and lasts for $400\kyr$  with a luminosity of approximately $\sci{6}{39}\ergs$. The secondary detaches as the separation grows to $280\Rsun$. Its RL is filled again when the star ascends the AGB. The star detaches when the separation grows to $1,200\Rsun$, but still nearly fills its RL. This results in significant MT rate through WRLOF mode that powers a ULX for $100\kyr$ with a luminosity of $\sim\sci{3}{39}\ergs$. Afterwards, the secondary forms a $0.79\Msun$ WD.

\subsection{Donors}

\begin{figure*}
    \centering
    \includegraphics[width=\textwidth]{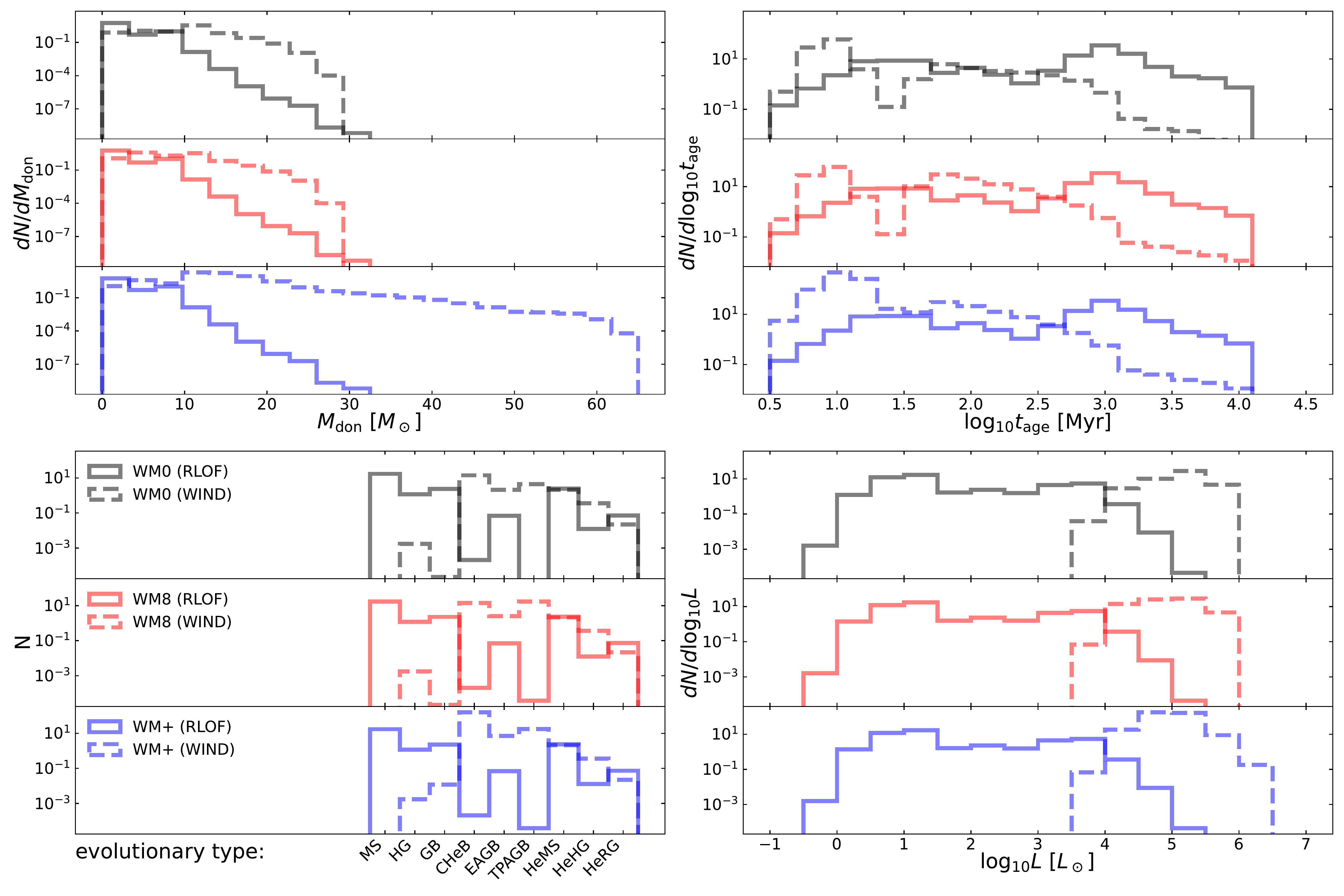}
    \caption{Distributions of ULX donor parameters: mass ($M_{\rm comp}$, upper left-hand panel), age since ZAMS at the moment of observation ($t_{\rm age}$, upper right-hand panel), evolutionary type (lower left-hand panel), and bolometric luminosity ($L$, lower right-hand panel). The maximum age of the ULXs was limited to $10\Gyr$ due to the duration of the simulation. Model names and accretion modes are defined in caption to Figure~\ref{fig:Lxdist}. Evolutionary types are: MS--main sequence; HG--Hertzsprung gap; RG--red giant; CHeB-Core Helium burning; EAGB--early asymptotic giant branch; TPAGB--thermal pulsing asymptotic giant branch; HeMS--Helium main sequence; HeHG--Helium Hertzsprung gap; HeRG--Helium red giant.}
    \label{fig:donordists}
\end{figure*}

Figure~\ref{fig:donordists} presents the parameter distributions of donors in the sample of wind-fed ULXs. Donors in RLOF ULXs were thoroughly investigated in our previous papers \citep[e.g.][]{Wiktorowicz1709} and are provided here only for reference.  

The masses of the donors in wind-fed systems are on average larger than in RLOF-fed ones and reach $30\Msun$ for WM0/8 models, or $\sim70\Msun$ for WM+ model. Wind-fed accretion does not suffer from dynamical instability for high mass-ratio systems, which is the case of RLOF stars with convective envelopes, and allows for $q\gtrsim3$. Meanwhile, the heavier the star, the shorter is its lifetime. This significantly limits the estimated number of ULXs with most massive donors and localizes them in star-forming regions or their immediate vicinity.

The evolutionary stage of wind-feeding donors is markedly different from those in RLOF ULXs, which are predominantly MS and He-rich stars. In wind-fed ULXs these are mostly CHeB and AGB stars that have significantly expanded due to nuclear evolution, and produce dense and slow stellar wind. Being more massive, donors in wind-fed systems are also on average much younger than in RLOF ULXs. However, those in orbit with NSs may be much older and the ULX phase can occur as late as $>1\Gyr$ after the ZAMS. We note that in a typical situation, the secondary becomes rejuvenated while accreting from stellar wind blown out from the evolving primary.

Donors in wind-fed ULXs are also more luminous than those in RLOF systems, mainly because of their higher masses and advanced evolutionary phase, with luminosities reaching $\sim10^6\Lsun$. These luminosities should be easily detectable at extragalactic distances, where the majority of ULXs are observed, and agrees with the fact that most of the donor candidates in ULXs are RSGs \citep{Heida1910}.

RSG donors are a minority in studies of ULX populations \citep[e.g. $\lesssim1\%$ in ][]{Wiktorowicz1709}, as we show also in this study ($\sim\sci{6.8}{-5}\%$). This did not produce a conflict with observations because donor candidates were only detected in a handful of ULXs and the observations were strongly biased toward detection of NIR objects. Additionally, ULXs are predominantly extragalactic objects, so low-mass, and therefore dim, stars, which theoretically dominate donors in RLOF-fed ULXs \citep[e.g.][]{Wiktorowicz1709}, are naturally excluded from detection. However, our study of wind-fed ULXs shows that in reality RSG might compose a significant fraction of ULXs, and in particular dominate those which are wind-fed ($\gtrsim88\%$; see Table~\ref{tab:dcos}). Moreover, $\gtrsim36\%$ of them may transfer mass through the WRLOF mode. Only in old stellar environments ($t_{\rm age}\gtrsim1\Gyr$) does the fraction of RSGs among ULXs peter out (Table~\ref{tab:results}) due to their short lifetimes. None of the ULXs with RSG donors will evolve into a merging DCO (Section~\ref{sec:dco}) and will instead typically form wide systems that are composed of a compact object and a carbon-oxygen WD.

According to our results, RSGs are expected to dominate the populations of wind-fed ULXs. This abundance of RSGs in our sample of synthetic ULXs suggests that the apparent connection between ULXs and RSGs in the observational data might be real.

\subsection{Double Compact Object Progenitors}\label{sec:dco}

\begin{deluxetable*}{llccccc}
    \tablewidth{\textwidth}
    \tablecaption{Double compact object progenitors among ULXs}
    \tablehead{ Model & MT Mode & & All & DCO & mDCO & WRLOF DCO}
    \startdata
WM0 & RLOF & all & $\mathbf{\sci{2.4}{1}}$ & \sci{7.2}{-3}\% & \sci{1.1}{-3}\% & -- \\
 &  & RSG & \sci{6.8}{-5}\% & \sci{1.7}{-5}\% & -- & -- \\\\
 & WIND & all & $\mathbf{\sci{2.3}{1}}$ & 46\% & 6.0\% & -- \\
 &  & RSG & 88\% & 39\% & -- & -- \\\\
WM8 & RLOF & all & $\mathbf{\sci{2.4}{1}}$ & \sci{7.2}{-3}\% & \sci{1.1}{-3}\% & -- \\
 &  & RSG & \sci{6.8}{-5}\% & \sci{1.7}{-5}\% & -- & -- \\\\
 & WIND & all & $\mathbf{\sci{3.7}{1}}$ & 32\% & 3.8\% & 3.2\% \\
 &  & RSG & 89\% & 27\% & -- & 3.0\% \\\\
WM+ & RLOF & all & $\mathbf{\sci{2.4}{1}}$ & \sci{7.2}{-3}\% & \sci{1.1}{-3}\% & -- \\
 &  & RSG & \sci{6.8}{-5}\% & \sci{1.7}{-5}\% & -- & -- \\\\
 & WIND & all & $\mathbf{\sci{1.9}{2}}$ & 15\% & 0.75\% & 14\% \\
 &  & RSG & 95\% & 14\% & -- & 14\% \\\\
    \enddata
    \tablecomments{Predicted number of all ULXs for the tested models and accretion modes (see caption of Table~\ref{tab:results}) with fractions of this number that fall into different categories: All--all systems; DCO--ULXs that are DCO progenitors; mDCO--ULXs that are progenitors of merging DCOs; WRLOF DCO--ULXs that transfer mass through WRLOF mode and are DCO progenitors. Calculated for a Milky Way-like galaxy with constant star-formation history. Rows labeled ``RSG'' present the sub-population of ULXs with RSG donors. There are no ULXs that are progenitors or merging DCOs and transfer mass through WRLOF mode. Zeros are marked as ``--'' for clarity.}
    \label{tab:dcos}
\end{deluxetable*}

ULXs are among the potential double compact object (DCO) progenitors \citep[e.g.][]{Mondal2001}. \cite{Heida1910} suggested that ULXs harboring RSGs, thus massive stars, are especially good candidates. However, as we have shown in Section~\ref{sec:er}, the majority of ULXs with RSG donors (and wind-fed ULXs in general) are expected to form either wide BH/NS-WD binaries or be disrupted. In this section, we analyze the evolutionary routes leading to the formation of DCOs from wind-fed ULXs, with a particular attention placed on merging ones (mDCOs; i.e. with time to merger, $t_{\rm merge}<10\Gyr$).

Up to $46\%$ of wind-fed ULXs (depending on the model) will form DCOs. Much fewer (up to $6\%$) represent progenitors of mDCOs (Table~\ref{tab:dcos}). Although the fractions of mDCO progenitors for tested models are different, all predict a similar expected number of these systems (about $1.4$ per Milky Way-like galaxy). This happens because all of them are nearly exclusively powered through the BHL accretion mode and none are powered through WRLOF, the prescription for which presents the only difference between models. We note that nearly all ($\gtrsim99\%$) of mDCO progenitors are wind-fed systems. Wind-powered systems can have much heavier donors ($M_{\rm ZAMS}\gtrsim8\Msun$; thus progenitors of compact objects) because the high mass-ratio ($\gtrsim3$) does not lead to CE, which is an expected outcome if the RL is filled.

The small fraction of mDCO progenitors among ULXs results mainly from the preference for NS-forming donors in these systems \footnote{Companion stars in ULXs are typically secondaries \citep[e.g][]{Wiktorowicz1709}; that is, less-massive stars on ZAMS. For a uniform mass ratio distribution, the secondaries mass is typically half the mass of the primary. Therefore, if the secondary is to have an initial mass larger than $\sim20\Msun$ (i.e. the lower limit for BH formation in isolation), then the primary should typically have an initial mass that is larger than about $40\Msun$. Due to the steepness of the IMF, such primary masses are rare.} and the resulting disruption of the system (due to the natal kick). Even if such a system survives a supernova explosion without being disrupted, or it has evolved without a RLOF/CE phase, it will be (on average) too wide for a merger to occur within $10\Gyr$, that is $t_{\rm merger}>10\Gyr$.

\begin{figure*}
    \centering
    \includegraphics[width=\textwidth]{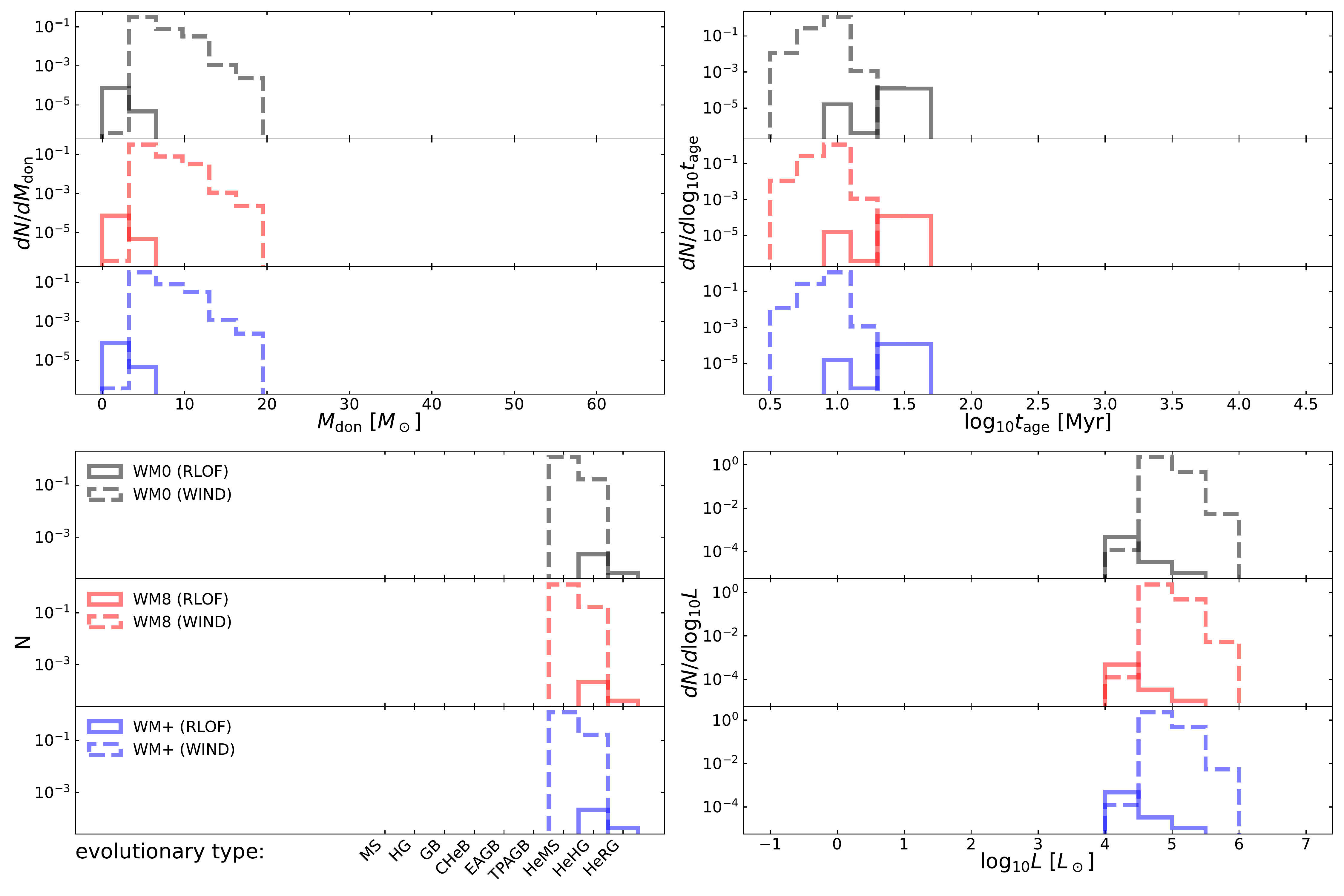}
    \caption{Parameter distributions of merging DCO progenitors among ULXs. Similar to Figure~\ref{fig:donordists} ($x$-axes kept the same for easier comparison). There are no progenitors that are fed through WRLOF accretion mode, therefore, distributions for all tested models are similar.}
    \label{fig:dcos}
\end{figure*}

The properties of mDCO progenitors are presented in Figure~\ref{fig:dcos}. The $x$-axis ranges remain the same as in the similar plots for the general distributions (Figure~\ref{fig:donordists}), so several distinct features are visible. Specifically, accretors are mostly BHs with a typical mass of $8$--$10\Msun$, whereas donors are evolved Helium-rich stars with a typical mass of $5$--$7\Msun$ and very high luminosities ($\log_{10}L/\Lsun\approx5$). These ULXs are typically very young objects with $t_{\rm age}\ll100\Myr$, so expected only in star-forming environments. Massive primaries and short orbital periods make the survival of the second SN more viable, but still most of these systems are disrupted \citep[e.g.][]{Wiktorowicz1911}

\begin{figure}
    \centering
    Merging DCO Progenitor Evolution\\
    \includegraphics[width=\columnwidth]{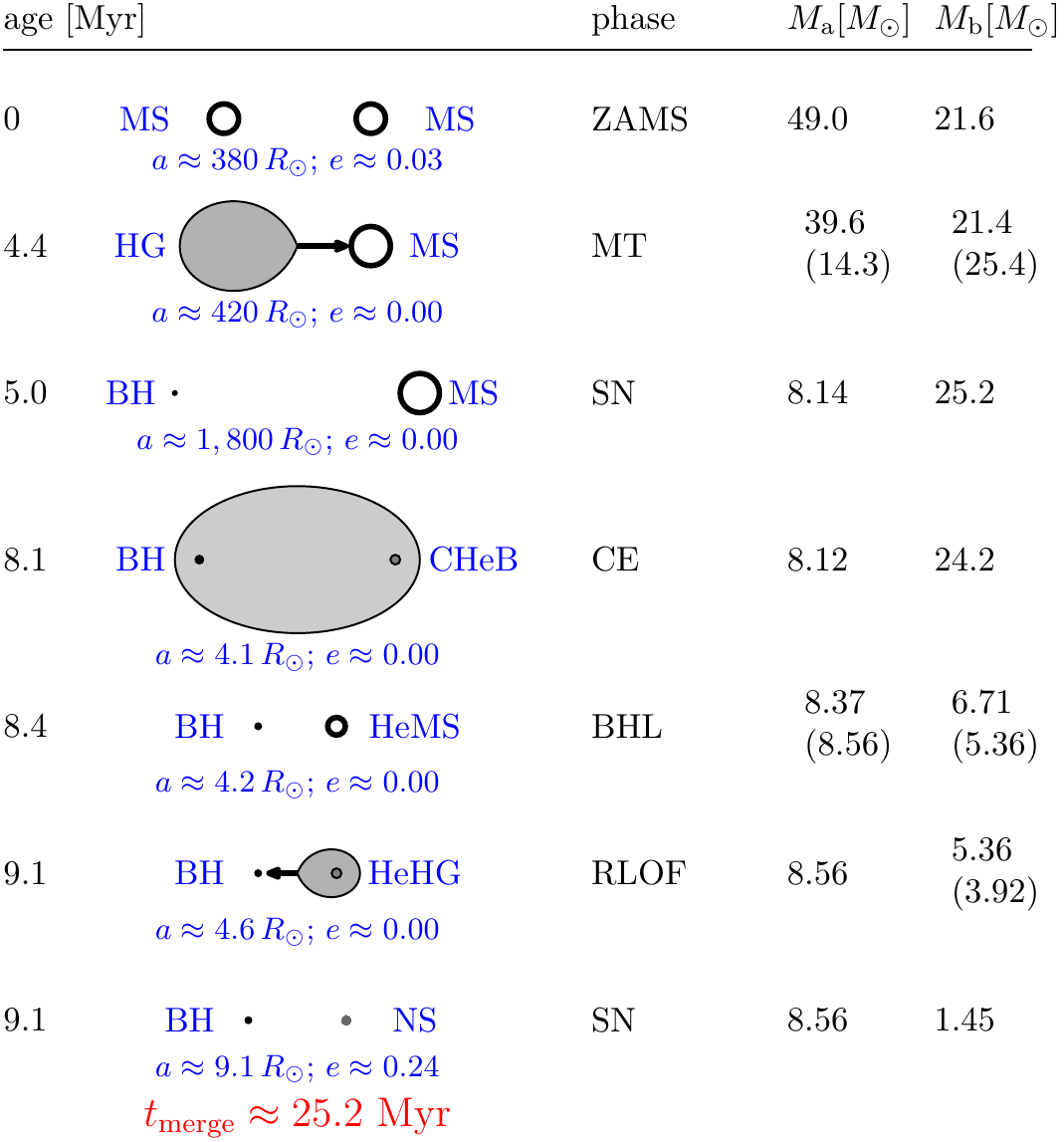}
    \caption{Same as Figure~\ref{fig:ts_NS}, but for a typical wind-fed ULX that is a progenitor of a merging DCO. Additional abbreviations (which are not explained in the captions to Figures~\ref{fig:ts_NS}, \ref{fig:ts_BH}, and \ref{fig:ts_RLOFWIND}), are for binary evolution: CE--common envelope binary; and for stellar evolution: HeMS--Helium main sequence; HeHG--Helium Hertzsprung gap.} 
    \label{fig:ts_mDCO}
\end{figure}

Progenitors of mDCOs among wind-fed ULXs are mainly composed of a BH accretor with $8$--$9\Msun$ and a $5$--$7\Msun$ evolved star that transfers mass through BHL mode. As stated earlier, there are no mDCO progenitors among ULXs powered through WRLOF mode because these are expanded post-MS stars on wide orbits and evolve typically into wide binaries with $t_{\rm merge}\gg10\Gyr$. In a typical case of a mDCO progenitor (Figure~\ref{fig:ts_mDCO}), both stars are initially relatively massive ($49.0\Msun$ and $21.6\Msun$), the separation is moderate ($a\approx 380\Rsun$), and the orbit is nearly circular ($e\approx0.03$). In about $4.4\Myr$, the primary evolves off the MS and commences a MT onto the secondary, which leads to a mass reversal. In the next half \Myr, the primary forms a BH with a mass of $8.14\Msun$, while the orbit expands to $1800\Rsun$. The secondary needs an additional $\sim3\Myr$ to become a giant and fill its RL starting the CE phase, which results in a significant shrinkage of the separation (down to about $4.1\Rsun$). With its hydrogen layers ripped off, the secondary becomes a Helium MS star with a stellar wind that is strong enough to power a ULX though a BHL mode (MT rate of $\sci{1.8}{-6}\msy$ out of which $\sim15\%$ is transferred to the vicinity of the BH, which is enough to power an isotropic ULX). After about $0.7\Myr$ it fills its RL and the MT mode switches to RLOF, which is much stronger and powers a hyper ULX (MT rate of $\sci{1.3}{-3}\msy$ and luminosity of $L_{\rm X}\approx10^{42}\ergs$). In a very short time ($800\yr$), the secondary explodes as a SN and forms a NS. The natal kick alters the orbit to a separation of $9.1\Rsun$ and eccentricity of $e\approx0.24$. The time to merger is $t_{\rm merge}=25.2\Myr$.

\section{Discussion}

\subsection{Observed Systems}

Although hundreds of ULXs have been detected so far \citep[e.g.][]{Walton1109, Swartz1111}, only a handful of them have viable donor candidates. The observed donors can help us to understand the nature of ULXs, but are difficult-to-observe objects. Except for potential ULXs, such as SS433 and GRS1915, all (non Be-X) ones are extragalactic objects. Therefore, only the most luminous donors are bright enough to be identified by contemporary telescopes. The situations may be even worse given that binary evolution studies predict that most of the companions in ULXs, especially those containing NS accretors, have low masses and are still on their MS \citep[$\lesssim3\Msun$;][]{Wiktorowicz1709}. Additionally, in most situations, the spatial coincidence of a star and an X-ray source is the only argument for physical connection. Many ULXs (especially with visible--i.e., massive--donor candidates) are localized in dense, star-forming regions, where more than one star can exist in the X-ray position error circle. Furthermore, the accretion disk light may exceed the optical luminosity of the donor or be hard to remove from the photometry \citep[e.g.][]{Grise1202,Sutton1411}. Therefore, it is usually only possible to look for optical counterparts only during the quiescent phases when the X-ray and optical emission from the disk is lowered. 

Many authors have reported the detection of stars spatially coinciding with the X-ray sources, naming them companion candidates. \citet{Motch1105} identified a potential optical counterpart to NGC7793 P13 with spectral features suggesting a supergiant luminosity class. \citet{Motch1410} estimated a mass of $18$--$23\Msun$ for this object, assuming that minimum light represents the stellar light. Their modeling of the disk optical emission caused by X-ray heating supported this assumption. The star was classified as B9Ia, which together with the orbital period of $\sim64$ days suggested the presence of a compact object with a mass less than $15\Msun$. Indeed, \citet{Furst1611} discovered that it is a magnetized NS. \citet{Heida1408} concentrated on NIR companions to nearby ($<10\mpc$) ULXs and detected $11$ potential NIR companions for $62$ ULXs. They claim to detect all luminous NIR sources (RSG candidates) in this volume, which leads us to conclude that only a small fraction of ULXs are possibly accompanied by RSGs. Using this sample, \citet{Heida1511} identified a potential counterpart to the ULX in NGC 253 as a RSG candidate. The authors suggest that the unusually high proper motion ($\sim66\kms$ larger than average for the Galaxy) strengthens their claim that the star and the ULXs are associated because such runaway RSGs are hard to explain otherwise. \citet{Heida1606} investigated five more ULXs from the \citet{Heida1408} sample and found that two of them are spatially coinciding with NIR sources that are compatible with being RSGs. 

The study by \citet{Heida1408} was later extended by \citet{Lopez1707} and \citet{Lopez2009} to a final sample of 113 ULXs within the distance of 10 Mpc (from the total population of 170 ULXs in this range). In summary, viable RSG candidate counterparts were found for five of these ULXs.

\citet{Heida1910} identified a RSG spatially coinciding with NGC 300 ULX-1, which harbours a NS accretor \citep{Carpano1805}. For such a high mass-ratio, the orbital modulations of the donor candidate are expected to be immeasurable, despite the source being the nearest ULX with a known accretor type \citep[$D=2.0\mpc$;][]{Dalcanton0907}.

\citet{Liu1311} detected a WR star in the ULXs in NGC 101, which provides MT onto the accretor through wind. The detection was based on spectroscopic analysis, so provides a stronger argument than just spatial coincidence.

For older environments the number of wind-fed ULXs is expected to drop significantly due to the short life span of supergiant stars, which constrain the majority of donors. Consequently, ULXs in old stellar populations (both RLOF- and wind-fed) are predicted to be dominated by NS accretors with low-mass donors \citep[$\lesssim3\Msun$;][]{Wiktorowicz1709}

\citet{elMellah1902} studied the importance of WRLOF on ULXs. They concentrated on two specific cases with mass ratios $q=15$ and $2$. They analyzed NGC 7793 P13 and M101 ULX-1, and showed that their calculations agree with the systems being wind-fed ULXs for the know orbital parameters.

\subsection{Wind Accretion Efficiency}\label{sec:accretion}

In our results, the typical accretion efficiency in wind-fed ULXs in the BHL mode is $1$--$2\%$, so only supergiants producing strong stellar wind can provide a MT rate large enough to power such a source. Meanwhile, in the WRLOF mode the accretion efficiency is much larger and typically reaches $30$--$50\%$. However, for WRLOF to operate, the wind must be slow inside the RL, so again supergiants are the most promising donors. We note that $50\%$ is a limit imposed on accretion efficiency in our reference model to avoid extrapolating the results of \citet[see also \citeauthor{Abate1304} \citeyear{Abate1304}]{Mohamed10}.

There are processes, which are not investigated in this study, that can decrease the efficiency of wind accretion. BHL is a simple model assuming the homogeneity of the accreted mass and a linear respective motion of the accretor and the gas. However, accretion in a binary system involves non-homogeneous winds, non-linear respective motions of the accretor and the wind-source (donor), as well as the presence of centrifugal forces. \citet{Theuns9606} performed a smooth particle hydrodynamic simulation of wind accretion in a situation where the orbital velocity is comparable to the wind velocity, and therefore cannot be neglected. They obtained mass-accretion rates $10$ times smaller than expected from the BHL prescription. In addition, \citet{Boffin9411} noted that an order of magnitude lower mass-accretion rates give better agreement with observation of enrichment in barium stars. However, in the majority of our results, the orbits of wind-fed ULXs are wide, and therefore the wind velocity is much higher than the orbital velocity and the above effects may not be so strong.

For high mass-accretion rates ($\dot{M}\gtrsim\dot{M}_{\rm Edd}$) the wake can become optically thick, which leads to instabilities that effectively reduce the accretion rate because of radiation momentum and energy deposition in the flow \citep{Taam9104}. Furthermore, \citet{Bermudez-Bustamante2004} showed that in AGB stars, most of the wind is lost through the L2 point and forms an excretion disk around the binary, effectively reducing the MT rate.

A detailed treatment of the accretion process may also lead to an increase of the wind accretion rates. For example, the formation of an accretion disk is an effect that has been ignored in the analytical solutions but which proved to be potentially important in hydrodynamical simulations, and can occur also in the BHL accretion mode \citep{Theuns9606}. The disk can help to reduce the angular momentum of the accretion flow and, consequently, increase the accretion rate. Furthermore, the accretion flow can be heated to a high temperature while the gravitational energy is being released, which can result in the production of highly-energetic radiation that can irradiate the donor, and thus increase its mass-loss rate. Unfortunately, hydrodynamical simulations cover only a small fraction of the parameter space and these results cannot be applied to a general case.

Similarly, the available hydrodynamical simulations of WRLOF accretion flows are scarce and the analytical fits, such as those of \citet{Abate1304} depend on restricting assumptions.In particular, the upper limit for the fraction of accreted mass can be higher than $50\%$ and a WRLOF may smoothly convert into a RLOF when the star fills its RL.

\subsection{Spatial Coincidence Probability for ULXs and RSGs}

\citet{Heida1511} investigated a ULX in NGC 273 (designated as J0047) that has a spatially coinciding RSG donor candidate. They calculated the probability that J0047 and the RSG are only located in the same place on the sky by chance as $<2.6\%$ through dividing the X-ray detection error circle ($95\%$ credibility) by the size of the observational field and multiplying this value by the number of RSGs (not-fainter than the RSG coinciding with J0047) in this field. This approach was later used for two other ULXs with spatially coinciding RSGs in \citet{Heida1606}.

We note that such an approach does not give a probability for spatial coincidence, which is a general concept for a population, but only gives an expected number of RSGs with this specific luminosity inside randomly located X-ray detection error circle for this particular field. It can be shown that by performing the same analysis for a (possibly imaginary) field with $\gtrsim100$ times more RSGs, in which case the procedure predicts a superposition chance above $100\%$.

The main problem of this method is that the object chosen for investigation (J0047) was not chosen at random, but due to the coinciding RSG. This is a relatively rare situation because only several such cases were found among hundreds of known ULXs. Only the analysis of the entire populations of ULXs and RSGs, or their representative samples, may allow us to perform reliable statistical estimates of the probability of having a ULX and a RSG in the same position on the sky. The situation may be further complicated by the fact that current observational surveys are biased toward bright NIR sources, effectively favoring RSGs while neglecting other unobserved stars that may be much better donor candidates.

Recently, \citet{Lopez2009} performed a combined analysis for 113 nearby ULXs, focusing on deriving the expected number of RSGs spatially coinciding with this sample. They obtained the expected number of $3.7$ such coincidences, which is of the same order as the number of ULX with RSG donor candidates (five systems) in their sample.

Despite the problems pointed out above with connecting ULXs to RSGs by spatial coincidence alone, we note that our theoretical analysis presented in this paper shows that a significant fraction of the ULXs ($\sim42$--$54\%$; Table~\ref{tab:dcos}) does posses RSG donors. Some of these sources may be hidden from our view because of misalignment of the emission beam and the line of sight.

\section{Summary}

In this study, we analyzed the population of wind-fed ULXs in the context of the entire ULX population. Using the population synthesis method and statistical analysis, we showed that wind-fed ULXs, which were mostly neglected in previous studies, constitute a significant fraction of all ULXs and in some environments may be a majority. Although our assumptions were rather optimistic, we prove that wind-fed ULXs cannot be neglected in research on ULXs, lest the study be systematically biased. Our sample contains a significant fraction of RSG companions, and thus supports the suggestion that the apparent superposition of some ULXs and RSGs results from coexistence as a binary. Although some of these systems will evolve into DCOs, none of the ULX systems harboring a RSG donor are viable progenitors of DCO mergers with $t_{\rm merge}<10\Gyr$, due to the large separations.

It should be stressed that the models of wind accretion and emission that are used in our study might not always be realistic, which could influence our results. This problem cannot be solved by extensive numerical simulations alone, abd requires us to obtain quantitative and qualitative observational data.
 
\acknowledgements

We are thankful to the anonymous referee who pointed out relevant points to make the paper more consistent. We also thank to thousands of volunteers who took part in the {\it Universe@Home} project\footnote{https://universeathome.pl/} and who provided their computers for simulation used in this study. G.W. is partly supported by the President’s International Fellowship Initiative (PIFI) of the Chinese Academy of Sciences under grant no. 2018PM0017 and by the Strategic Priority Research Program of the Chinese Academy of Science Multi-wave band Gravitational Wave Universe (grant No. XDB23040000). This work is partly supported by the National Natural Science Foundation of China (grant No. 11690024, 11873056, and 11991052) and the National Key Program for Science and Technology Research and Development (grant No. 2016YFA0400704). K.B. and G.W. acknowledge the NCN grant MAESTRO 2018/30/A/ST9/00050. This work is partly supported by the National Key Program for Science and Technology Research and Development (grant No. 2016YFA0400704), and the National Natural Science Foundation of China (grant No. 11690024 and 11873056). J.P.L. was supported in part by a grant from the French Space Agency CNES. K.B. acknowledges support from the Polish National Science Center (NCN) grant Maestro (2018/30/A/ST9/00050). 

\bibliographystyle{apj}
\bibliography{ms}

\end{document}